\documentclass[showpacs,aps,twocolumn]{revtex4}

\usepackage{bm}
\usepackage{amsmath}
\usepackage{graphicx}
\usepackage{subfigure}
\usepackage[usenames,dvipsnames]{color}
\definecolor{darkblue}{RGB}{0,0,196}
\usepackage[colorlinks=true,linkcolor=darkblue,citecolor=darkblue,urlcolor=darkblue]{hyperref}

\usepackage{xcolor}

\usepackage{setspace}
\usepackage{footmisc}
\usepackage[makeroom]{cancel}
\usepackage{comment}
\usepackage{lineno}

\usepackage{graphicx}
\usepackage{amsmath,bbm}
\usepackage{amssymb,bm}
\usepackage{yfonts}

\usepackage{bm}
\usepackage{amsmath}
\usepackage{graphicx}
\usepackage{subfigure}
\usepackage[usenames,dvipsnames]{color}
\definecolor{darkblue}{RGB}{0,0,196}
\usepackage[colorlinks=true,linkcolor=darkblue,citecolor=darkblue,urlcolor=darkblue]{hyperref}
\usepackage{setspace}
\usepackage{hyperref}
\usepackage{footmisc}
\usepackage[makeroom]{cancel}
\usepackage{comment}
\usepackage{lineno}

\def\be{\begin{equation}}
\def\ee{\end{equation}}
\def\ba{\begin{eqnarray}}
\def\ea{\end{eqnarray}}

\usepackage{graphicx}
\usepackage{amsmath,bbm}
\usepackage{amssymb,bm}
\usepackage{yfonts}

\newcommand \rpi{\pi^{+}+\pi^{-}}
\newcommand \rk{\rm{K^{+}+K^{-}}}

\newcommand \rlambda{\Lambda}

\newcommand\tch{T_{\rm{ch}}}

\newcommand \sq{\sqrt{s}}

\begin{document}
\title{A Baseline Study of the Event-shape and Multiplicity Dependence of Chemical Freeze-out Parameters in Proton-Proton Collisions at $\sqrt{s}$ = 13 TeV Using~PYTHIA8}
\author{Rutuparna Rath$^{a}$}
\author{Arvind Khuntia$^{b}$}
\author{Sushanta Tripathy$^{a}$}
\author{Raghunath Sahoo$^{a}$\footnote{Corresponding author: $Raghunath.Sahoo@cern.ch$}}
\affiliation{$^{a}$Department of Physics, Indian Institute of Technology Indore, Simrol, Indore 453552, India}
\affiliation{$^{b}$The H. Niewodniczanski Institute of Nuclear Physics, Polish Academy of Sciences, PL-31342 Krakow, Poland}

\begin{abstract}
The event-shape and multiplicity dependence of the chemical freeze-out temperature ($T_{\text{ch}}$), freeze-out radius ($R$), and strangeness saturation factor ($\gamma_{s}$) are obtained by studying the particle yields from the PYTHIA8 Monte Carlo event generator in proton-proton (pp) collisions at the centre-of-mass $\sqrt{s}$ = 13 TeV. Spherocity is one of the transverse event-shape techniques to distinguish jetty and isotropic events in high-energy collisions and helps in looking into various observables in a more differential manner. In this study, spherocity~classes are divided into three categories, namely (i) spherocity integrated, (ii) isotropic, and~(iii) jetty. The~chemical freeze-out parameters are extracted using a statistical thermal model as a function of the spherocity class and charged particle multiplicity in the canonical, strangeness canonical, and grand canonical ensembles. A clear observation of the multiplicity and spherocity class dependence of $T_{\text{ch}}$, $R$, and $\gamma_{s}$ is observed. A final state multiplicity, $N_{\rm ch}\geq$ 30 in the forward multiplicity acceptance of the ALICE detector appears to be a thermodynamic limit, where the freeze-out parameters become almost independent of the ensembles. This~study plays an important role in understanding the particle production mechanism in high-multiplicity pp collisions at the Large Hadron Collider (LHC) energies in view of a finite hadronic phase lifetime in small systems.

\end{abstract}
 
\date{\today}
\maketitle

\section{Introduction}
Quark Gluon Plasma (QGP), a~deconfined state of quarks and gluons, is believed to be produced in heavy-ion collisions at the Relativistic Heavy-Ion Collider (RHIC) and the Large Hadron Collider (LHC). However, the~observations of QGP signatures like the strangeness enhancement~\cite{ALICE:2017jyt} and double ridge structure~\cite{Khachatryan:2016txc} in high-multiplicity proton-proton (pp) collisions at the LHC indicate the possible formation of QGP droplets in pp collisions. These discoveries have important consequences on whether to use the pp collisions as a baseline to understand a medium formation in heavy-ion collisions. Thus, a~closer look at the underlying physics mechanisms in pp collisions has become very essential. Many of the QGP-like behaviours have been successfully explained by phenomena such as multi-partonic interactions, string fragmentation, color~reconnection, rope hadronization, etc., which are incorporated in the PYTHIA8 Monte Carlo event generator~\cite{pythia8html}. Unlike the lower collision energies, where pp has been used as a reference measurement to study heavy-ion collisions, the~pp collisions at the LHC energies have brought up new challenges and opportunities in terms of their high-multiplicity environment to study many emergent phenomena.
 In this direction, one uses the recently introduced transverse spherocity to separate jetty and isotropic events in pp collisions, as~the production dynamics for both are different. When the jetty events involve high transverse momentum ($p_{\rm T}$) phenomena and are described by perturbative Quantum Chromodynamics (pQCD), the~isotropic events are mostly dominated by soft-physics (low-$p_{\rm T}$). It would be interesting to look into the dependence of chemical freeze-out parameters on event-shape and multiplicity. In~order to do that, we use a statistical thermal model in different tunes of ensembles and employ the method of transverse spherocity, which is more effective in discriminating multi-jet topologies~\cite{Cuautle:2014yda} than some of the other techniques like sphericity~\cite{Banfi:2010xy} and thrust~\cite{Kar:2018uno}. As~such a study has not been performed in any of the LHC experiments, the~present analysis using the PYTHIA8 event
 generator would be useful in performing a baseline study and bringing out necessary physics~messages.

The thermal statistical model has been quite successful in describing the composition of stable particles in high-energy pp, p-Pb, Xe-Xe, and Pb-Pb collisions at the LHC~\cite{Sharma:2018jqf,Rath:2019aas}. In~general, the~grand canonical ensemble is used for heavy-ion collisions as the volume of the produced system is large enough so that it holds the relation $VT^{3}>1$, where $V$ and $T$ are the system volume and temperature, respectively. However, for small collision systems, where the number of particles is small, one has to explicitly conserve the charges of the QCD, namely baryon~number (B), strangeness (S), and electric charge (Q) in the small volume. Such a canonical treatment of particle production results in the suppression of particles carrying non-zero quantum numbers as it has to be created in pairs. In~this framework, the yields of non-strange particles in small systems are expected to be very similar to those observed in large collision systems, whereas the strangeness production is expected to be suppressed in smaller systems. However, a recent multiplicity-dependent study of the production of strange and multi-strange particles relative to pions in pp collisions at the LHC showed an enhancement, and for high-multiplicity pp collisions, the~obtained values are similar to Pb-Pb collisions~\cite{ALICE:2017jyt}. This observation has raised one of the key questions about whether the systems created in high-multiplicity pp collisions have reached a thermodynamical limit where both the canonical and grand canonical ensembles are equivalent. This was addressed in~\cite{Sharma:2018jqf,Rath:2019aas}, where it was indeed observed that the chemical freeze-out parameters obtained in high-multiplicity pp collisions are similar to the peripheral Pb-Pb collisions. To~address this behavior, one can also use the strangeness canonical ensemble where the exact conservation of strangeness is required while the baryon and charge content is treated grand canonically. Studying the chemical freeze-out parameters (CFO) such as the chemical freeze-out temperature ($T_{\rm ch}$), strangeness saturation factor ($\gamma_s$), and freeze-out radius ($R$) using multiplicity alone could be an incomplete study as in each multiplicity class, we can have jetty, as well as isotropic events. In~view of the observation of
heavy-ion-like features in high-multiplicity pp collisions~\cite{ALICE:2017jyt,Khachatryan:2016txc}, the~studies of the event topology dependence of particle production have been an
emerging area of research at the LHC energies~\cite{Acharya:2019mzb,Khatun:2019dml,Tripathy:2019blo,Khuntia:2018qox,Cuautle:2015kra}. Recently, a~finite hadronic phase lifetime was also reported using the event multiplicity dependence of ALICE data~\cite{Sahu:2019tch,Acharya:2019qge}. This hints at looking into the event-shape dependence of chemical freeze-out parameters in pp collisions at the LHC energies. Before~an experimental study is made, it is worth looking into the possibilities using a widely used pQCD-based event generator like PYTHIA8~\cite{Sjostrand:2006za}, which has been successful in describing many observed features in pp collisions. One should note that PYTHIA8 does not have inherent thermalization, while~for a study on the CFO properties, one needs to have the thermalization of a system. However, as~reported in~\cite{Ortiz:2013yxa}, the~color reconnection (CR) mechanism along with the multipartonic interactions (MPIs) in PYTHIA8 produces the properties that arise from the thermalization of a system such as radial flow and the mass dependent rise of mean transverse momentum. In~the PYTHIA model, a~single string connecting two partons follows the movement of the partonic endpoints, and this movement gives a common boost to the string fragments (final state hadrons). With~CR along with MPI, two partons from independent hard scatterings can reconnect, and they increase the transverse boost. This microscopic treatment of final state particle production is quite successful in explaining the similar features that arise from a macroscopic picture via the hydrodynamical description of high-energy collisions. Thus, it is apparent that the PYTHIA8 model with MPI and CR has a plausible ability to produce the features of thermalization, and this allows us to look for the CFO properties using simulated data from PYTHIA8. The~present study should serve as a baseline work in this direction exploring the CFO properties in pp collisions and their event topology and multiplicity dependence. This need arises because of the different dynamics of particle production in jetty and isotropic~events.

To separate the jetty and isotropic events from the average-shaped events, one should look into the geometrical shape of events using event-shape observables. Transverse spherocity as one of the event-shape observables has given a new direction for understanding the underlying events in pp collisions to have further differential study along with event multiplicity. Recently, the~kinetic freeze-out scenario and system thermodynamics were studied using event-shape and event multiplicity in pp collisions at the centre-of-mass $\sq$ = 13 TeV using PYTHIA8~\cite{Tripathy:2019blo,Khuntia:2018qox}. It would be interesting to look into the spherocity dependence of chemical freeze-out parameters, which would provide a complete picture of the produced fireball and hadronic phase in high-multiplicity pp collisions. In~this work, we perform a double differential study of chemical freeze-out parameters using spherocity and final state charged particle multiplicity in pp collisions at $\sq$ = 13 TeV using~PYTHIA8.

\begin{figure}[ht!]
\includegraphics[scale=0.4]{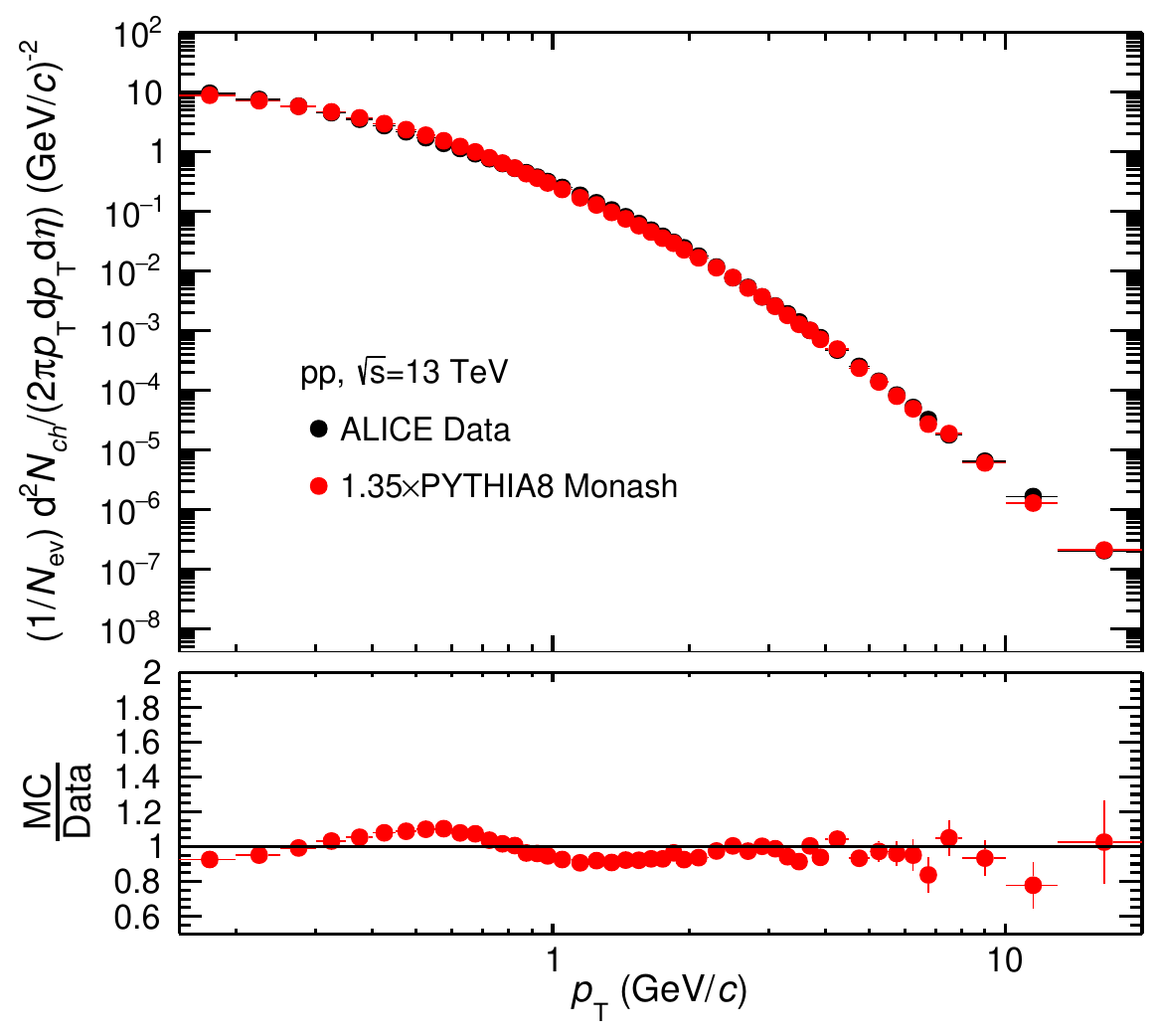}
\caption[]{(Color Online) Upper panel: comparison of charged particle $p_{\rm T}$ spectra in pp collisions at $\sqrt s$~=~13~TeV between ALICE data~\cite{Adam:2015pza} and Monte Carlo (MC) simulation, which are used for this analysis. Lower~panel: the ratio between scaled simulated data and experimental data.}
\label{datVsPH}
\end{figure}

\begin{figure}[ht!]
\includegraphics[scale=0.4]{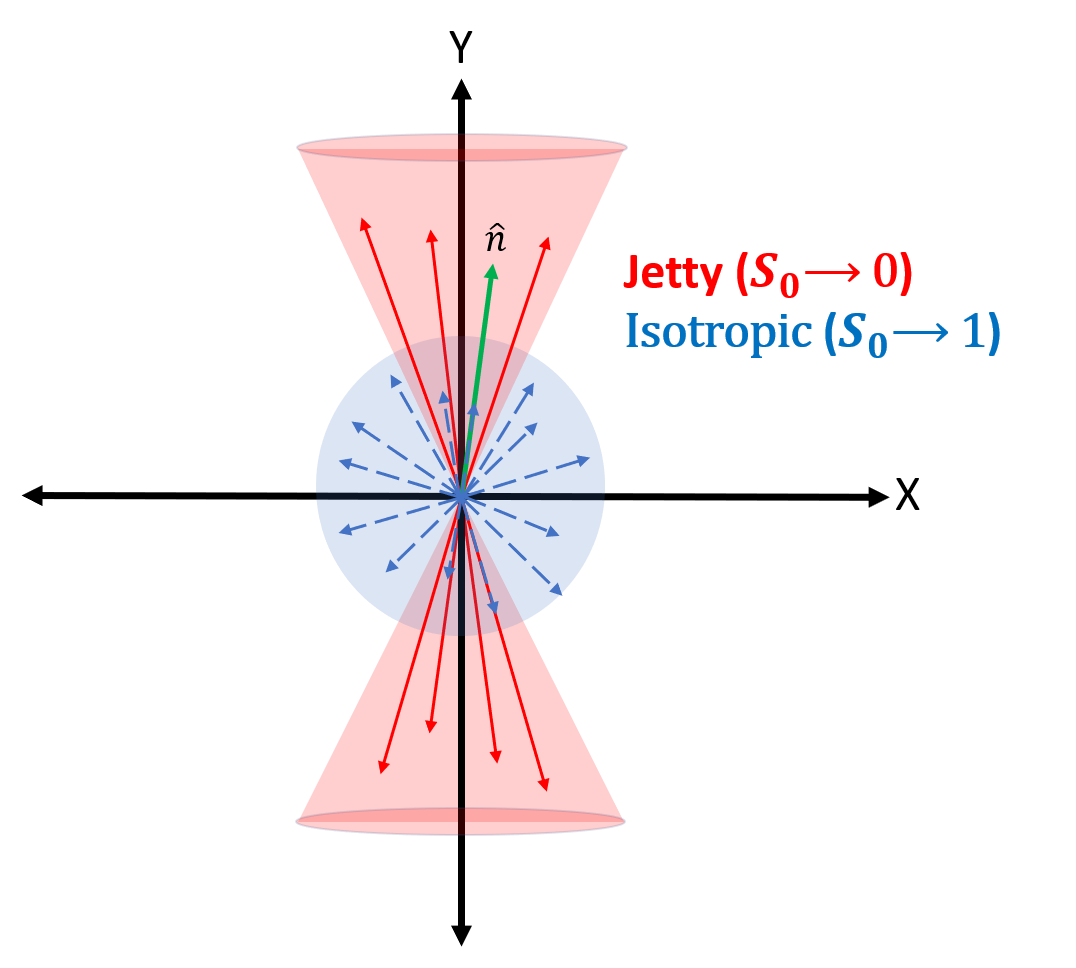}
\caption[]{(Color Online) Jetty and isotropic events in the transverse plane.}
\label{sp_cart}
\end{figure}

This paper is organized as follows.~The~description of event generation in PYTHIA8, the analysis~methodology, and the definition of spherocity are given in Section~\ref{ev_ana}. In~Section~\ref{sec:1}, we discuss in detail the different ensembles used in this study. The~results obtained using different ensembles for different spherocity classes are discussed in Section~\ref{sec:2}. Finally, in Section~\ref{sec:3}, we conclude with a summary of the important~results.
\section{Event Generation and Analysis Methodology}
\label{ev_ana}
For this analysis, we used the PYTHIA8 to simulate ultra-relativistic proton-proton collisions. It is a blend of many models and theories relevant for physics like parton distributions, hard~and soft interactions, initial and final-state parton showers, fragmentation, multipartonic interactions, color~reconnection, and decay~\cite{Sjostrand:2006za}. We used Version 8.235 of PYTHIA, which includes MPI, which is crucial to explain the underlying events and multiplicity distributions, along~with other observables. Furthermore, as~explained in the previous section, this version includes flow-like patterns in terms of color reconnection~\cite{Ortiz:2013yxa}. The~detailed description of the physics processes in PYTHIA8.235 can be found in~\cite{pythia8html}. We implemented the inelastic, non-diffractive component of the total cross-section for all soft QCD processes using the switch (SoftQCD: all = on). Around 250 million events were generated at $ \sqrt{s}=13~\mathrm{TeV} $ with Monash 2013 tune (Tune:14)~\cite{Skands:2014pea}. We used the MPI based color reconnection scheme (ColorReconnection:mode(0)) as it describes the ALICE charged particle spectra in pp collisions at $\sqrt s$ = 13 TeV, as shown in Figure~\ref{datVsPH}, the~details of which are discussed in Section~\ref{sec:2}. One should note that for our analysis, the~minimum bias events are those events where no selection of charged particle multiplicity and spherocity is applied. For~the generated events, all the resonances were allowed to decay except the ones used in our study (HadronLevel:Decay = on). In~our analysis, the event selection criteria are such that only those events were chosen that had at least five tracks (charged particles at the mid-rapidity $|\eta| < 0.8$), and the charged particle multiplicities ($N_{\rm ch}$) were chosen in the acceptance of the V0 subdetector with a pseudorapidity range of V0A ($2.8<\eta<5.1$) and V0C ($-3.7<\eta<-1.7$)~\cite{Abelev:2014ffa} to match the experimental conditions of the ALICE detector. The~events generated using these cuts were then further divided into ten multiplicity (V0M) classes, each class containing 10\% of the total events. Table~\ref{tab:V0M} shows all ten charged particle multiplicity~classes.

\begin{table}[h]
\caption{V0M multiplicity classes and the corresponding charged particle multiplicities.}
\centering 
\scalebox{0.8}
{
\begin{tabular}{|c|c|c|c|c|c|c|c|c|c|c|} 
\hline 
V0M class & 1 & 2 & 3 & 4 & 5 & 6 & 7 & 8 & 9 & 10 \\
\hline 
$N_{\rm ch}$ &50-140 & 42-49 & 36-41 & 31-35 & 27-30 & 23-26 & 19-22 &  15-18 & 10-14 & 0-9\\
\hline
\end{tabular}
}
\label{tab:V0M}
\end{table}
It is worth noting here that although there are many tunes of physics processes in PYTHIA used to explain observable
specific physics findings, our aim in this paper is to see the event-shape and multiplicity dependence of CFO parameters taking the best 
suited~tunes. 

Transverse spherocity is defined for a unit vector $\hat{n} (n_{\rm T},0)$, where $n_{\rm T}$ is the transverse component of the unit vector $\hat n$, that minimizes the following quantity~\cite{Cuautle:2014yda}:
\begin{eqnarray}
S_{0} = \frac{\pi^{2}}{4} \bigg(\frac{\Sigma_{i}~|\vec p_{{\rm T}_{i}}\times\hat{n}|}{\Sigma_{i}~p_{{\rm T}_{i}}}\bigg)^{2},
\end{eqnarray}
where $p_{\rm T_i}$ is the transverse momentum of the $i^{\rm th}$ particle.
The distinct configuration of events if they are isotropic or jetty in the transverse plane is coupled to the extreme limits of spherocity, which varies from zero to one. In~the spherocity distribution, the~events limiting towards one are isotropic in nature, while towards zero are jetty. The~jetty events are the consequence of hard processes, while the isotropic events are soft processes. The~depiction of jetty and isotropic events is shown in Figure~\ref{sp_cart}. The~spherocity distribution is selected in the pseudo-rapidity range of $|\eta|<0.8$, and all events have a minimum constraint of five charged particles with $p_{\rm{T}}~ $ $>$ 0.15 GeV/$c$. In~the spherocity distribution, the jetty events are those having $0\leq S_{0}<0.29$ with the lowest 20 percent of total events, and the isotropic events are those having $0.64<S_{0}\leq1$ with the highest 20~percent of the total~events. 

\section{Use of ensembles in thermal model}
\label{sec:1}

In this study, ($\rpi$)/2, ($\rk$)/2, (${\rm p} + {\bar{\rm p}}$)/2, $\phi$, and $(\Lambda+\bar{\Lambda})/2$ yields at mid-rapidity obtained from PYTHIA8 are used to extract the CFO parameters in THERMUS~\cite{Wheaton:2004qb} based on the canonical, strangeness canonical, and grand canonical ensembles. For~convenience, here onwards, ($\rpi$)/2, ($\rk$)/2, (${\rm p} + {\bar{\rm p}}$)/2, and $(\Lambda+\bar{\Lambda})/2$ are denoted as $\pi$, K, p, and $\Lambda$, respectively. Now we discuss these three ensembles in~details.

\subsection{Grand canonical ensemble (GCE)}
 The grand canonical ensemble (GCE) is generally used in applications to heavy-ion collisions. In~this ensemble, the energy and number of particles along with their quantum numbers are enforced by conservation laws on average by temperature and chemical potential. The~partition function for a system with $N$ hadrons within a volume $V$ with temperature $T$ and~chemical potential $\mu$ is given by

\begin{eqnarray}
\begin{split}
\ln Z^{GCE}(T,V,\{\mu_i\}) = \sum_i{g_iV\over\left(2\pi\right)^3}\int d^3p\\
\ln\left(1\pm exp(-\left(E_i-\mu_i\right)/T)\right)^{\pm1}.
 \end{split}                      
\end{eqnarray}

Here, $E_i^{2} = p_i^{2}+m_i^{2}$ is the total energy ($p_i$ being the momentum of a particle of mass $m_i$), $g_{i}$ is the spin-isospin degeneracy factor, and $\mu_{i}$ is the chemical potential of the $i^{\rm{th}}$-species. $T$ is the temperature, and the ``+`` and ``--'' signs in the distribution functions refer to bosons and fermions, respectively. Here, the~individual particle numbers are not conserved, while the quantum numbers $B$, $S$, and~$Q$ are conserved. For the~$i^{\rm{th}}$-hadron, the~chemical potential is given by 

\begin{eqnarray}
\mu_i &=& B_i\mu_B + S_i\mu_S + Q_i\mu_Q.
\end{eqnarray}

Here, B$_i$, S$_i$, and Q$_i$ are the baryon number, strangeness, and electric charge of the $i^{\rm{th}}$-hadron and $\mu_{B}$, $\mu_{S}$ and $\mu_{Q}$ are the corresponding chemical potentials for these conserved quantum numbers, respectively. Applying the Boltzmann approximation to the partition function \cite{Wheaton:2004qb}, we get, 
\begin{equation}
\ln Z^{GCE}(T, V, \{\mu_i\}) = \sum_i \frac{g_i V}{(2\pi)^3} \int d^3p\exp\left( -\frac{E_i-\mu_i}{T} \right).\\
\end{equation}

Hence, the~particle multiplicity in GCE is given by,
\begin{equation}
N_i^{\rm GCE} = \frac{g_i V}{(2\pi)^3} \int  \exp \left( -\frac{E_i-\mu_{i}}{T}\right)d^3p.
\end{equation}

At the LHC energies,~$\mu$ $\simeq$ 0, and the expression for particle multiplicities becomes,
\begin{equation}
N_i^{\rm GCE} = \frac{g_i V}{(2\pi)^3} \int \exp \left( -\frac{E_i}{T}\right)d^3p.
\end{equation}
 
The particle yields measured by detectors in ultra-relativistic collisions also include the feed down from the heavier hadrons and resonances. This has a significant contribution for lighter particles like pions. Thus, the final yield is given by 
\begin{equation}
N_i^{\rm GCE}(\mathrm{total}) = N_i^{\rm GCE} + \sum_j {\rm Br}(j\rightarrow i) N_j^{\rm GCE}.
\end{equation}

Here, Br$(j\rightarrow i)$ is the number of the $i^{\rm{th}}$-species into which a single 
particle of species $j$ decays.

\subsection{Strangeness canonical ensemble (SCE)}
The strangeness canonical ensemble (SCE), also called the mixed canonical ensemble in THERMUS terminology, requires the exact conservation of the strangeness quantum number~\cite{Toneev:2004kg}, while the baryon and charge content is treated grand canonically. Thus, the~partition function is given by 
\begin{equation}
Z_{\rm SCE} = \frac{1}{(2\pi)}
\int_0^{2\pi} \exp(-iS\varphi)
Z_{\rm GCE}(T,\mu_B,\lambda_S)d\varphi.
\end{equation}

Here, we shall only consider the case where overall strangeness is zero, $S=0$. The~chemical potential of the $i^{\rm{th}}$-hadron is given by
\begin{eqnarray}
\mu_i &=& B_i\mu_B+ Q_i\mu_Q,
\end{eqnarray}

and the fugacity factor is replaced by
\begin{eqnarray}
\lambda_S = \exp(i\varphi).
\end{eqnarray}
 
As previously done for the case of GCE, the~decays of resonances have to be added to the final yields. Thus, the~expression for particle multiplicity is given by
\begin{equation}
N_i^{\rm SCE}(\mathrm{total}) = N_i^{\rm SCE} + \sum_j {\rm Br}(j\rightarrow i) N_j^{\rm SCE} .
\end{equation}

\subsection{Canonical Ensemble}
In the canonical ensemble (CE), the~conservation of quantum numbers corresponding to $B$, $S$, and $Q$ is exactly enforced. 
 The~partition function is given by
\begin{equation}
\begin{split}
Z^{CE} = &\frac{1}{(2\pi)^3}
\int_0^{2\pi} d\alpha exp(-iB\alpha)
\int_0^{2\pi} d\psi exp(-iQ\psi) \\
&\int_0^{2\pi} exp(-iS\varphi)
Z_{GCE}(T,\lambda_B,\lambda_Q,\lambda_S) d\varphi.
\end{split}
\end{equation}

Here, the fugacity factor is replaced by
\begin{equation}
\lambda_B = exp(i\alpha),\quad \lambda_Q = exp(i\psi), \quad \lambda_S = exp(i\varphi)   .
\end{equation}
The feed downs from resonances have to be added to the final yields similar to the case of GCE,
\begin{equation}
N_i^{\rm CE}(\mathrm{total}) = N_i^{\rm CE} + \sum_j {\rm Br}(j\rightarrow i) N_j^{\rm CE} .
\end{equation}

These ensembles are implemented in the THERMUS program, and we refer to~\cite{Wheaton:2004qb} for more detailed steps and their implementation in~THERMUS.

The number density, energy density, and pressure of each hadron species in the Boltzmann approximation within the canonical ensemble differ from the grand canonical ensemble by a multiplicative factor with all the chemical potentials set to zero, as discussed in~\cite{Wheaton:2004qb}. This multiplicative correction factor depends on the thermal parameters and the quantum numbers of the particle. For~large systems, one approaches the grand canonical ensemble, while the extension of thermal model to elementary collisions requires an additional parameter known as the strangeness saturation factor, $\gamma_{s}$, which accounts for the deviation from chemical equilibrium in the strangeness sector. The~possibility for the incomplete strangeness equilibrium is achieved by multiplying $\gamma_{s}^{|s_{i}|}$ by the thermal distribution function by replacing~\cite{Becattini:2003wp}:
\begin{equation}
\exp \left( -\frac{E_i-\mu_{i}}{T}\right) \rightarrow \exp \left( -\frac{E_i-\mu_{i}}{T}\right) \gamma_{s}^{|s_{i}|}.
\end{equation}

Here, $|s_{i}|$ is the number of valence strange quarks and anti-quarks in the $i^{\rm{th}}$-species.

Further, it should be noted here that in the case of high-energy collisions, one expects a scaling of the boost-invariant Bjorken expansion 
in the mid-rapidity plateau~\cite{Bjorken:1982qr}, which results in full phase space particle ratios being equivalent to the experimentally measured particle rapidity density ratios at the mid-rapidity~\cite{Cleymans:1999st,Broniowski:2001we}, 
\begin{equation}
\frac{dN_i/dy}{dN_j/dy} = \frac{N_i^0}{N_j^0},
 \end{equation}
 where $N_i^0 (N_j^0)$ is the particle yield of hadron species $i(j)$ in the rest frame of the fireball.~The~exact conservation of quantum numbers is expected in the $4\pi$ phase space and may not be the case for one unit of rapidity. However, the~deviations in the case of one unit of rapidity 
may not be large for the high-multiplicity events, but one could also check applicability at low-multiplicity. Hence, it may be good enough to assume conservation and study the particle yields at mid-rapidity, as~the particle
multiplicities are the highest at mid-rapidity. It is also shown that the strangeness canonical ensemble gives the best description of 
ALICE data for the rapidity window $\Delta{y}$ = 1.33 $\pm$ 0.28 using THERMUS~\cite{Vislavicius:2016rwi}.

Keeping this in mind and considering the above three ensembles, we proceed with our calculation of the CFO parameters in different multiplicity and spherocity classes in the next section using the identified particle yields obtained from the PYTHIA8.


\section{Results and discussion}
\label{sec:2}

To check the compatibility of the simulated data from PYTHIA8 with the experimental data, we compare the charged particle $p_{\rm T}$ spectra for pp collisions at $\sqrt{s}$ = 13 TeV from ALICE data~\cite{Adam:2015pza} and from the PYTHIA8 simulation used for this analysis. The~comparison is shown in Figure~\ref{datVsPH}. The~lower panel shows the ratio of the $p_{\rm T}$ spectra predictions from PYTHIA8 to experimental data. In~order to see the agreement of the spectral shapes, we used an arbitrary scaling factor (1.35) to scale the simulated data. The scaling factor used is to check the matching of the spectral shape, and it bears no physical significance. We found that the scaled simulated data agree with the spectral shape from the experimental data within (10--20)\% at low-$p_{\rm T}$ and consistent to unity for intermediate and high-$p_{\rm T}$. 

The grand canonical, strangeness canonical, and canonical ensembles are considered for pp at $\sq$~=~13~TeV using the PHYTHIA8 Monash tune for three spherocity classes, namely spherocity integrated, isotropic, and jetty events. We considered $\pi$, K, p, $\phi$, and $\Lambda$ yields in THERMUS for our study. We assigned 10\% uncertainties to the individual particle yields as a conservative higher limit in this analysis; as~usually, the~experimental uncertainties are of this order. The~simulated results in high-multiplicity pp collisions at $\sq$ = 13 TeV for different spherocity classes allow us to access very high-multiplicity events in small collision systems and further investigate the applicability of these three ensembles.
Looking~into the particle-antiparticle symmetry at the LHC energies, the~net baryon number and strangeness numbers are set to zero. Hence, the canonical ensemble differs from the grand canonical ensemble by a multiplicative factor as discussed in~\cite{Wheaton:2004qb}. However, in~the thermodynamic limit, they must be~equivalent.

\begin{figure}[!ht]
\begin{center}
\includegraphics[width=8cm,height=8cm]{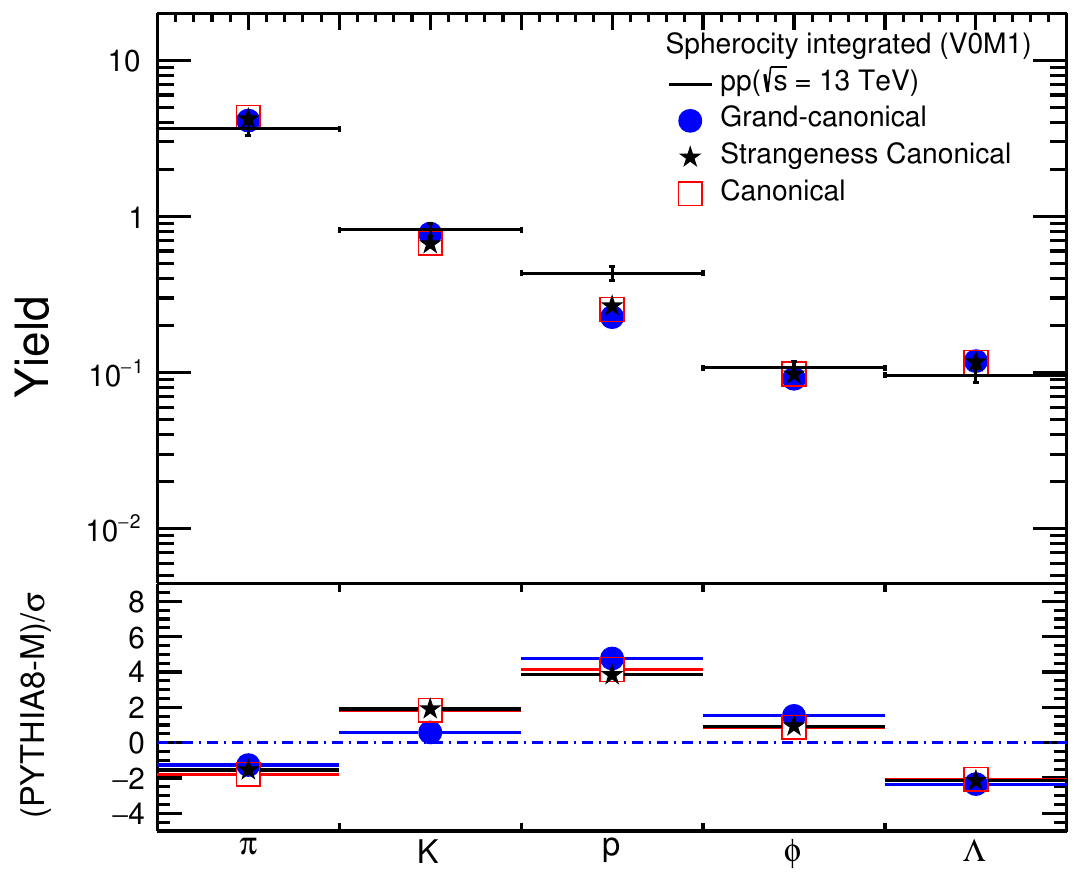}
\caption{(Color online) The upper panel shows the comparison of the identified particle yields in pp collisions at $\sq$ = 13 TeV for spherocity integrated events at the highest multiplicity class with the thermal model using the grand canonical, strangeness canonical, and canonical ensembles. The~lower panel shows the deviation from the fit~results.}
\label{fit:pp13:spint}
\end{center}
\end{figure}

\begin{figure}[!]
\begin{center}
\includegraphics[width=8cm,height=8cm]{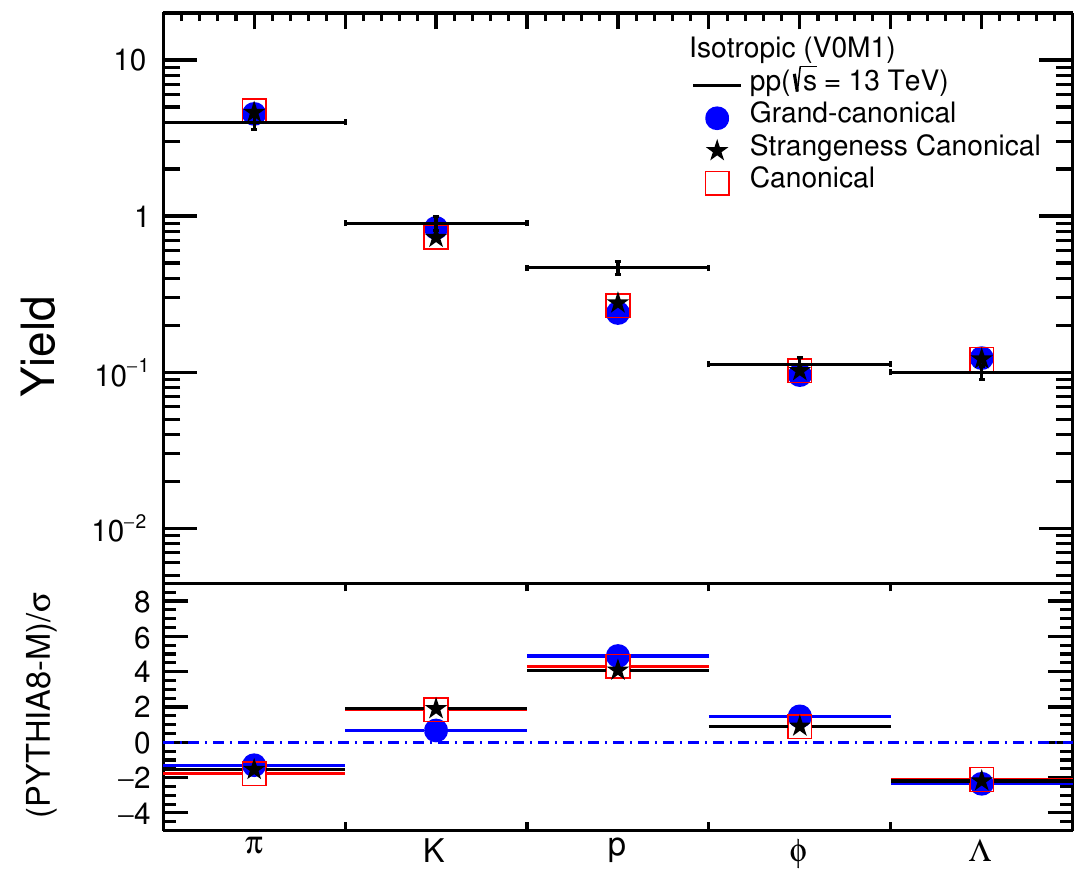}
\caption{(Color online) The upper panel shows the comparison of identified particle yields in pp collisions at $\sq$ = 13 TeV for isotropic events at the highest multiplicity class with the thermal model using the grand canonical, strangeness canonical, and canonical ensembles. The~lower panel shows the deviation from~the fit.}
\label{fit:pp13:iso}
\end{center}
\end{figure}

\begin{figure}[!ht]
\begin{center}
\includegraphics[width=8cm,height=8cm]{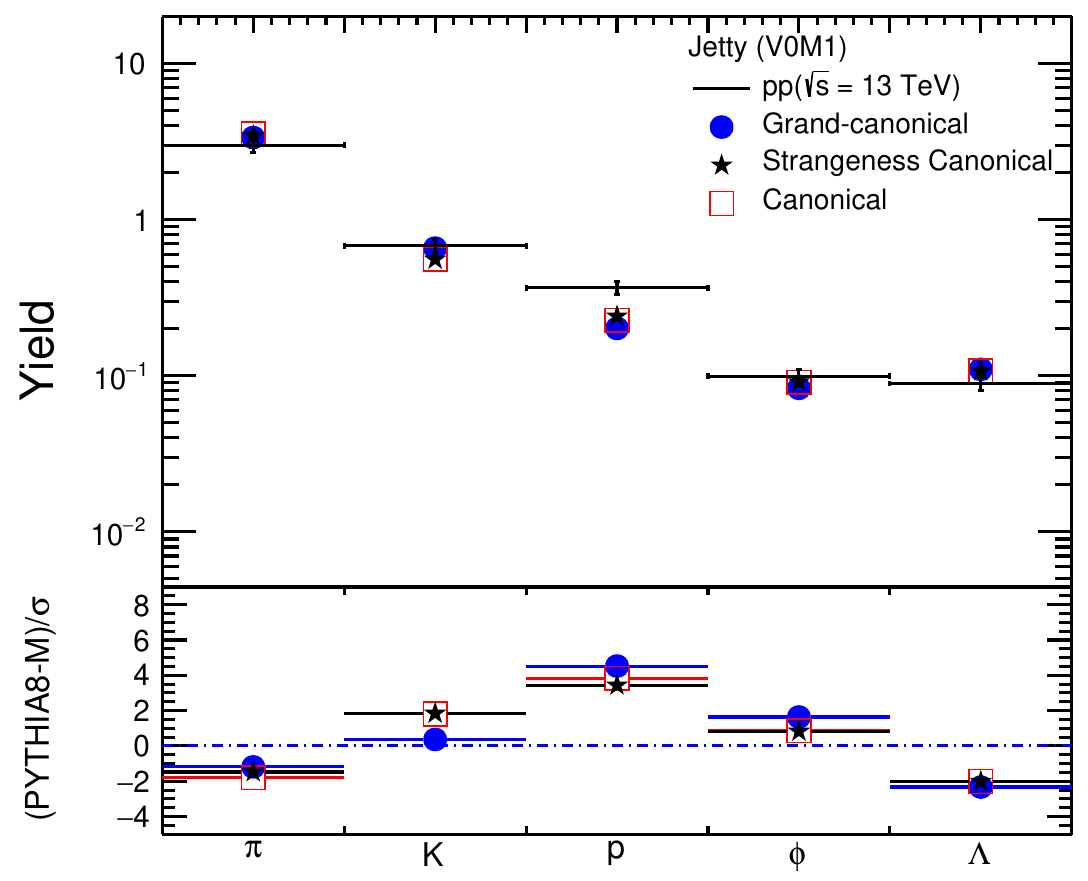}
\caption{ (Color online) The upper panel shows the comparison of the identified particle yields in pp collisions at $\sq$ = 13 TeV for jetty events at the highest multiplicity class with the thermal model using grand canonical, strangeness canonical and canonical ensemble. The~lower panel shows the deviation from fit~results.}
\label{fit:pp13:jetty}
\end{center}
\end{figure}


The particle yields in pp collisions are fitted using THERMUS~\cite{Wheaton:2004qb} for the grand canonical, strangeness canonical, and canonical ensembles with three free parameters, namely $\tch$, $\gamma_{s}$, and $R$. In~the case of SCE, we have an additional free parameter, known as the strangeness canonical radius, which is taken as the fireball radius (i.e.,~$R= R_c$) ~\cite{Wheaton:2004qb}. The~comparison between the model and data for the spherocity integrated class for the highest multiplicity class is shown in Figure~\ref{fit:pp13:spint}. In~the top panel, the~black bar shows the simulated data points obtained using PYTHIA8. The~blue solid circles are obtained for the grand canonical ensemble. The black solid stars are for the strangeness canonical ensemble. The red rectangular boxes are for the canonical ensemble.
The bottom panel shows the deviation from the fit, which is defined as
\begin{equation}
\text{Deviation ~from ~Fit} = \frac {\text{PYTHIA8 $-$ Model}}{\sigma}.
\end{equation}

Here, $\sigma$ are the uncertainties in particle yields as estimated using PYTHIA8. 
Similarly, the~comparison of the particle yields in the thermal model and data for the isotropic and jetty events for the highest multiplicity class are shown in Figures~\ref{fit:pp13:iso} and \ref{fit:pp13:jetty}, respectively. The~fitting of the thermal model to the simulated data points is performed as a function of charged particle multiplicity and spherocity classes for all the ensembles. The~goodness of fit and $\chi^{2}$/NDF, where NDF is the number of degrees-of-freedom, for different multiplicity and spherocity classes in pp collisions at $\sq$~=~13~TeV for different ensembles are given in Tables~\ref{tab:pp:iso}--\ref{tab:pp:spi}. The~obtained $\chi^{2}$/NDF is better for the higher charged particle multiplicity classes. The~extracted thermodynamic parameters are shown in Figures~\ref{fig:radius}--\ref{fig:gammas} as a function of charged particle multiplicity for three different spherocity classes. Although~one expects isotropic events to be better described by statistical thermal models, within~the obtained sensitivity of $\chi^{2}$/NDF between jetty and isotropic events, it is difficult to comment on. In~view of this, it would be nice to perform a similar study when experimental data are available in the near future, as~there is no inbuilt thermalization feature in PYTHIA8, except~the fact that MPI with CR mimics a thermalization feature---an outcome that gives a description of flow-like features seen in high-multiplicity pp events~\cite{Ortiz:2013yxa}. {{It is known that at the LHC energies, the proton-to-pion ratio is not well explained by the statistical hadron resonance gas model, which introduces some degree of uncertainty in the estimated thermal parameters}~\cite{Abelev:2012wca}{. This is reflected in the observed} $\chi^{2}${/NDF.} {With~baryon and anti-baryon annihilation channels being switched on in hydrodynamic models like VISHNU, the~proton-to-pion ratio at the LHC was well explained}~\cite{vishnu}. In~this direction, this work attempts to look for any possible dependency of the thermal parameters on event-shapes in hadronic collisions at the LHC, rather than emphasizing the exact estimation of the thermal parameters, which would be performed on the experimental~data.

\begin{figure}[!ht]
\begin{center}
\includegraphics[width=8.5cm,height=6. cm]{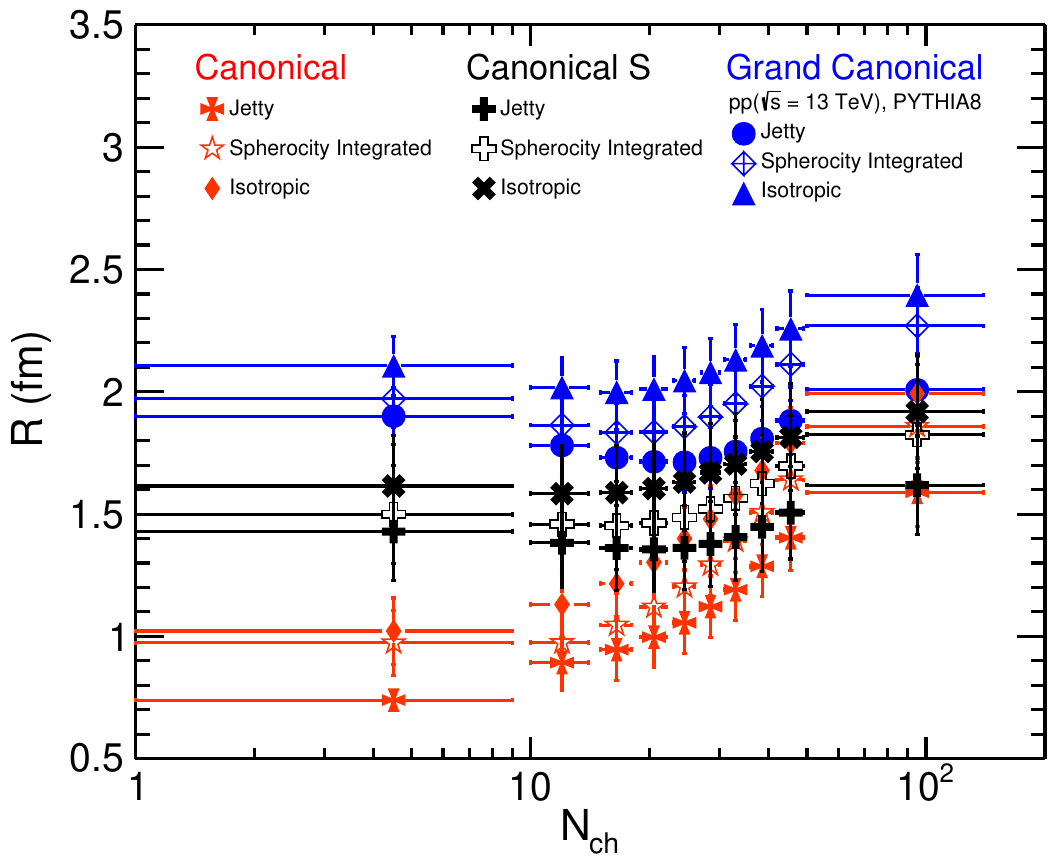}
\caption{ The extracted freeze-out radius as a function of different spherocity classes (shown by different markers) and charged particle multiplicity ($N_{ch}$) for the canonical, strangeness (S) canonical, and grand canonical ensembles. The red and black points are obtained using the canonical ensemble and the strangeness canonical ensemble, respectively, whereas the blue points are for the grand canonical ensemble.}
\label{fig:radius}
\end{center}
\end{figure}

\begin{figure}[!ht]
\begin{center}
\includegraphics[width=8.5cm,height=6.cm]{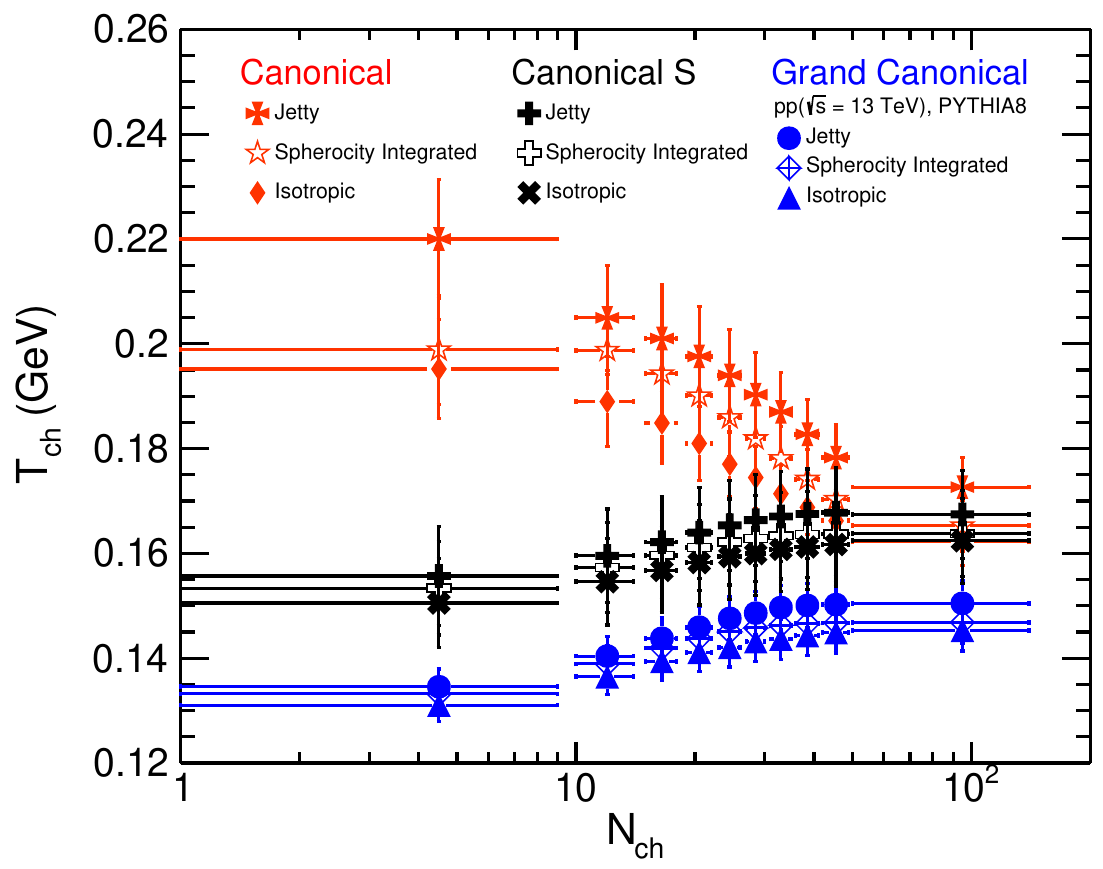}
\caption{ The chemical freeze-out temperature as a function of charged particle multiplicity ($N_{ch}$) for the canonical, strangeness canonical, and grand canonical ensembles. The~red and black points are obtained using the canonical ensemble and the strangeness canonical ensemble, respectively, whereas the blue points are for the grand canonical~ensemble.}
\label{fig:tch}
\end{center}
\end{figure}

In Figure~\ref{fig:radius}, we show the fireball radius $R$ as a function of the charged particle multiplicity for different spherocity classes in pp collisions at $\sq$ = 13 TeV by considering the grand canonical, strangeness canonical, and canonical ensembles. The~freeze-out radius increases with the charged particle multiplicity for all three ensembles. However, the~radius obtained for the canonical ensemble is lower compared to other two ensembles for a lower charged particle multiplicity. The~radius in the canonical ensemble approaches the strangeness canonical ensemble for the above charged particle multiplicity around 30. For~the strangeness canonical ensemble, the~value of the radius lies between the canonical and grand canonical ensembles. 
For~isotropic events, the~obtained radius is higher compared to the jetty and the spherocity integrated ones, as seen by ALICE~\cite{Acharya:2019idg}. It is observed that the fireball radius is lower for the jetty-like events. However, for~the high-multiplicity classes, the~behavior of the fireball radius is similar for all spherocity classes within the uncertainties. In~all three ensembles, the~observed behavior is~similar.

In Figure~\ref{fig:tch}, the~chemical freeze-out temperature $\tch$ is shown as a function of the charged particle multiplicity for different spherocity classes in pp collisions at $\sq$ = 13 TeV by taking the grand canonical, strangeness canonical, and canonical ensembles. In~general, the~freeze-out temperature obtained for the canonical ensemble is higher compared to the GCE and SCE. The~trend of the freeze-out temperature shows a slight increase with the charged particle multiplicity for the GCE and SCE, while the CE shows a decreasing behavior as a function of charged particle multiplicity. Beyond~a charged particle multiplicity of around 30, the~freeze-out temperature for the CE is similar to the SCE within~the uncertainties.

\begin{figure}[!ht]
\begin{center}
\includegraphics[width=8.5cm,height=6. cm]{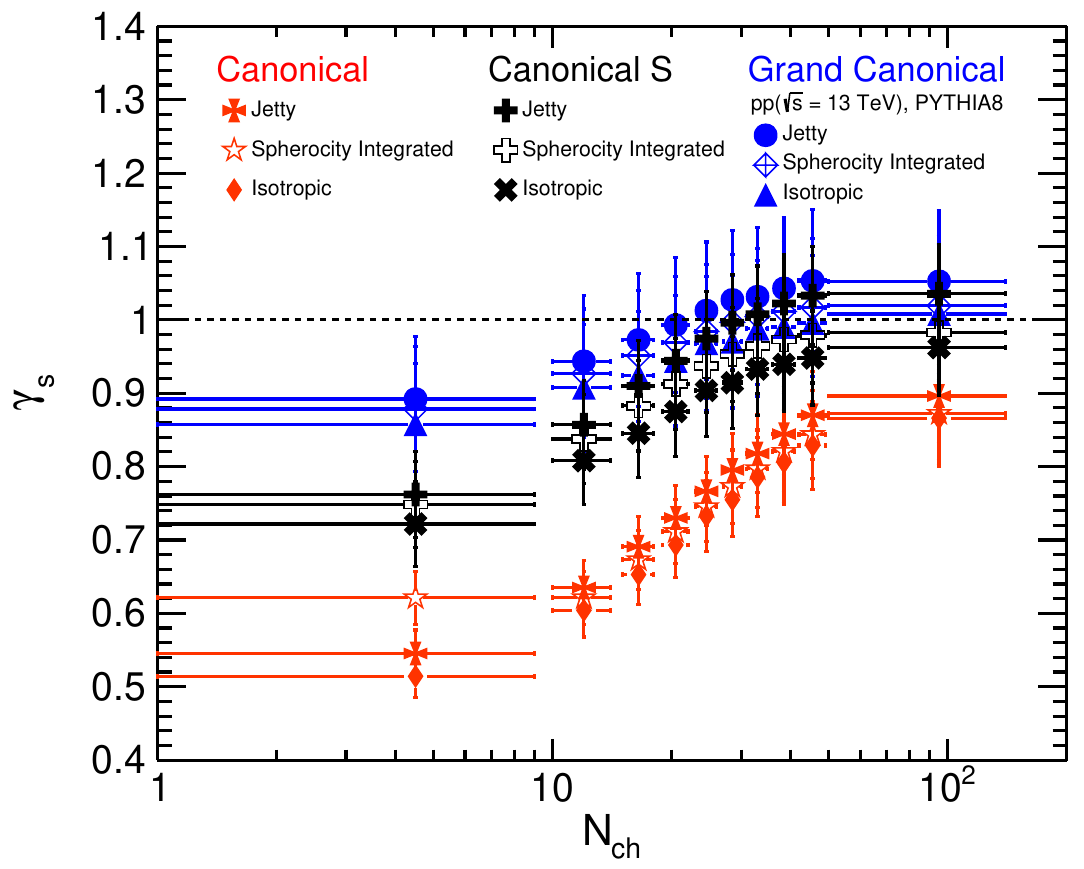}
\caption{ (Color online) The strangeness saturation factor ($\gamma_{s}$) as a function of the charge particle multiplicity ($N_{ch}$) and spherocity classes (shown by different markers) for the canonical, strangeness canonical, and grand canonical ensembles. The~red and black points are obtained using the canonical ensemble and the strangeness canonical ensemble, respectively, whereas the blue points are for the grand canonical~ensemble.}
\label{fig:gammas}
\end{center}
\end{figure}


\begin{table}[!]
\caption{$\chi^{2}$/NDF of the fits for isotropic events in different multiplicity classes for pp collisions at $\sqrt{s}$~=~13~TeV.}
\label{tab:pp:iso}
\begin{tabular}{ |p{2cm}||p{2cm}|p{2cm}|p{2cm}| }
 \hline
 \multicolumn{4}{|c|}{Isotropic Events } \\
 \hline
 Multiplicity Classes& Strangeness Canonical& Canonical& Grand Canonical\\
 \hline
 V0M1 & 28.215/2&30.284/2 & 33.681/2\\
 V0M2&   27.052/2& 30.092/2& 33.711/2\\
 V0M3&   26.706/2 & 30.432/2& 34.209/2\\
 V0M4&   26.636/2& 31.062/2& 34.935/2\\
 V0M5&   26.441/2& 31.463/2 &26.019/2\\
 V0M6& 27.021/2& 32.629/2 & 36.2792\\
 V0M7& 27.916/2& 33.829/2 & 36.88/2\\
 V0M8& 29.762/2& 35.686/2 & 38.361/2\\
 V0M9& 33.494/2& 38.961/2& 41.379/2\\
 V0M10& 41.636/2& 44.328/2& 46.803/2\\
 \hline
 \end{tabular}
\end{table}

\begin{table}[!]
\caption{$\chi^{2}$/NDF of the fits for jetty events in different multiplicity classes for pp collisions at $\sqrt{s}$ = 13~TeV.}
\label{tab:pp:jetty}
\begin{tabular}{ |p{2cm}||p{2cm}|p{2cm}|p{2cm}| }
 \hline
 \multicolumn{4}{|c|}{Jetty Events } \\
 \hline
 Multiplicity Classes& Strangeness Canonical& Canonical& Grand Canonical\\
 \hline
V0M1 & 22.138/2&25.978/2 & 30.078/2\\
 V0M2&   21.295/2& 26.429/2& 30.891/2 \\
 V0M3&   20.498/2& 26.408/2& 30.889/2\\
 V0M4&   20.407/2& 26.848/2& 31.085/2\\
 V0M5&   21.226/2&28.074/2 &32.199/2\\
 V0M6& 22.319/2& 29.223/2 & 32.961/2\\
 V0M7& 24.464/2& 31.153/2 & 34.477/2\\
 V0M8& 27.174/2& 33.297/2& 36.325/2\\
 V0M9& 32.124/2& 36.997/2 & 39.691/2\\
 V0M10& 41.109/2& 43.634/2& 46.007/2\\
 \hline
 \end{tabular}
\end{table}

\begin{table}[!]
\caption{$\chi^{2}$/NDF of the fits for spherocity integrated events in different multiplicity classes for pp collisions at $\sqrt{s}$ = 13~TeV.}
\label{tab:pp:spi}
\begin{tabular}{ |p{2cm}||p{2cm}|p{2cm}|p{2cm}| }
 \hline
 \multicolumn{4}{|c|}{Spherocity Integrated Events } \\
 \hline
 Multiplicity Classes& Strangeness Canonical& Canonical& Grand Canonical\\
 \hline
V0M1 & 26.255/2&28.850/2 & 32.449/2\\
 V0M2&   24.871/2& 28.679/2& 32.636/2 \\
 V0M3&   24.022/2 & 28.698/2& 32.745/2\\
 V0M4&   23.565/2& 29.015/2& 33.006/2\\
 V0M5&   23.564/2& 29.608/2&33.436/2\\
 V0M6& 24.117/2& 30.515/2& 34.024/2\\
 V0M7& 25.598/2& 32.055/2 & 35.167/2\\
 V0M8& 27.996/2& 34.113/2 & 36.912/2\\
 V0M9& 32.396/2& 37.473/2 & 39.976/2\\
 V0M10& 41.168/2& 40.147/2& 46.069/2\\
 \hline
 \end{tabular}
\end{table}
As mentioned in the previous section, the~strangeness saturation factor $\gamma_{s}$ is responsible for the degree of deviation from the strangeness chemical equilibrium in the strangeness sector. Thus, it is important to understand the strange particle production in smaller collision systems as a function of charged particle multiplicity and spherocity classes. Figure~\ref{fig:gammas} shows the strangeness saturation factor $\gamma_{s}$ as a function of the charged particle multiplicity for different spherocity classes in pp collisions at $\sq$ = 13 TeV by considering the grand canonical, strangeness canonical, and canonical ensembles. The~strangeness saturation factor $\gamma_{s}$ increases with the multiplicity in pp collisions and reaches unity for the higher charged particle multiplicity for the grand canonical and strangeness canonical ensembles. This indicates the strangeness chemical equilibrium for high-multiplicity pp collisions at $\sq$ = 13 TeV. For~higher multiplicity classes, $\gamma_{s}$ is similar for the grand canonical and strangeness canonical ensembles. However, when we study the yields with exact conservation of all three quantum numbers, i.e, for~the canonical ensemble,~$\gamma_{s}$ is below one for the higher charged particle multiplicity classes. In~the case of the spherocity dependence study of $\gamma_{s}$, it is higher for the jetty events followed by the spherocity integrated and isotropic events, although the separation is not very significant considering the~uncertainties.

As we observed a significant dependence of the CFO parameters on spherocity and multiplicity classes, to~compliment/confront our results, it would be interesting to look into the particle yields in pp collisions at $\sq$ = 13 TeV once the spherocity-dependent experimental data points are~available.

\section{Summary and conclusion}
\label{sec:3}

Recent heavy-ion-like properties such as enhanced production of strange particles with respect to pions~\cite{ALICE:2017jyt}, the degree of collectivity~\cite{Khuntia:2018znt}, the hardening of particle spectra with multiplicity~\cite{Dash:2018cjh,Tripathy:2018ehz}, etc., are seen in high-multiplicity events in proton-proton (pp) collisions at the LHC energies and have drawn considerable interest in the research community. In~order to understand the production dynamics of particles in high-multiplicity pp collisions, an~event-shape-dependent study becomes inevitable. Further to study the event topology dependence of the CFO properties, as~thermalization in the medium is a basic requirement, the PYTHIA8 model with multi-partonic interaction and the color reconnection mechanism offers the possibility to make such a study. Thus, in~this manuscript, using simulated data from PYTHIA8, we extracted CFO parameters such as the chemical freeze-out radius ($R$), temperature ($\tch$), and strangeness saturation factor ($\gamma_s$) by studying the particle yields in a thermal model for pp at centre-of-mass $\sq$ = 13 TeV in different multiplicity and spherocity classes with a primary focus on the dependence of the CFO parameters with the geometrical shape of an event. Here, three types of ensembles, namely, grand canonical, strangeness canonical, and canonical, are used to look into their applicability in small collision systems. This study was performed by considering $\pi$, K, p, $\phi$, and $\rlambda$ yields in THERMUS. The~important findings are summarized below: 

\begin{itemize}
\item The freeze-out radius increases with charged particle multiplicity, and the highest value is observed for the grand canonical ensemble followed by the strangeness canonical and canonical~ensembles.

\item For higher charged particle multiplicity, the~radii obtained in the canonical ensemble are similar to the strangeness canonical~values.

\item As the process of isotropization takes more time, the~radii are expected to be higher for isotropic events compared to jetty and spherocity integrated events. However, in~our study, we do not see such behavior within the~uncertainties. 

\item The chemical freeze-out temperature $\tch$ increases with the charged particle multiplicity for the grand canonical and strangeness canonical ensembles, whereas the canonical ensemble shows a reverse~trend.

\item As seen in the case of the radius parameter, the~chemical freeze-out temperature is similar for the strangeness canonical and canonical ensembles at higher charged particle~multiplicities.

\item The strangeness saturation factor, $\gamma_{s}$, shows a clear evolution as a function of charged particle multiplicity and reaches the value of one for the GCE and SCE, which indicates complete strangeness chemical~equilibrium. 

\item The strangeness saturation factor, $\gamma_{s}$, is higher for the grand canonical ensemble and minimum for the canonical~ensemble.

\item A final state multiplicity in the V0M acceptance of around $N_{\rm ch}\geq$ 30 appears to be a thermodynamic limit, where the freeze-out parameters become almost independent of the ensembles. This goes in line with earlier observations~\cite{Sharma:2018jqf,Thakur:2017kpv}. 

\item Although the~spherocity dependence of the CFO parameters shows a separation for different geometrical event shapes, considering the uncertainties, they are found to be~similar.

\end{itemize}

 In this study, the grand canonical and strangeness canonical ensembles converge for the higher charged particle multiplicity classes. For~high-multiplicity pp collisions, $\gamma_{s}$ reaches unity, which shows a full strangeness chemical equilibrium. It will be more interesting to look into the high-multiplicity pp events in different spherocity classes in the future when experimental data are available to get more information about the strangeness chemical equilibrium and to understand the particle production mechanism in smaller collision~systems.

\section*{Acknowledgement}
The authors acknowledge the financial support from ALICE Project No. SR/MF/PS-01/2014-IITI(G) of the Department of Science $\&$ Technology, Government of India. R.R. and S.T. acknowledge the financial support by the DST-INSPIRE program of the Government of India. R.S. acknowledges the financial support from DAE-BRNS Project No. 58/14/29/2019-BRNS of the Government of India. The~authors acknowledge fruitful discussions with Jean~Cleymans and Anton Adronic. Further, we appreciate the help of Ashish Bisht, MSc student of IIT Indore, for PYTHIA data generation. Fruitful discussions with Suman Deb are highly appreciated. The~authors further acknowledge the usage of the resources of the LHC grid computing facility of VECC, Kolkata.

\end{document}